\documentclass[epjST]{svjour}

\usepackage{graphicx}
\usepackage{amssymb,amsmath}

\def\mb{\mu_{\mathrm{B}} }
\def\x{\textit{x}}
\def\y{\textit{y}}
\def\z{\textit{z}}
\def\Na{${}^{23}\mathrm{Na}$ }

\def\kink{e_{\vec{k}}}
\def\kinkt{e_{2\vec{k}_0-\vec{k}}}

\begin{document}

\title{Critical velocity of antiferromagnetic spin-1 Bose-Einstein condensates at finite temperature}

\author{Gergely Szirmai\inst{1}\fnmsep\thanks{\email{szirmai.gergely@wigner.mta.hu}}
}

\institute{Institute for Solid State Physics and Optics, Wigner Research Centre, Hungarian Academy of Sciences, H-1525 Budapest P.O. Box 49, Hungary
}

\abstract{
We study the instability of a moving spinor Bose-Einstein condensate when the speed of flow reaches the critical velocity. This we identify on the basis of Landau's criterion, i.e. the velocity above which some elementary excitation energy becomes negative.  We show that the first-to-become unstable excitations are spin-carrying quasiparticles. We also discuss the temperature dependence of the critical velocity in a more advanced mean-field approximation.
}

\maketitle

\section{Introduction}
\label{intro}

The nature of superfluidity of Bose systems is the result of a remarkable interplay between Bose--Einstein condensation and particle-particle interaction. The complex phenomena of superfluids had been explained by  a remarkably simple two fluid model \cite{PitaevskiiBook}. One of the key concepts is the critical velocity, that is, the limit velocity of an obstacle  immersed in the superfluid and moving against it without exciting density waves, therefore the motion is exempt from damping. In Landau's original argumentation for the critical velocity only Galilean invariance and energy and momentum conservation were taken into account besides a specific dispersion relation of the elementary excitations \cite{PitaevskiiBook}. Later it was shown by Bogoliubov that such dispersion relation is indeed the consequence of Bose-Einstein condensation and the critical velocity is the speed of sound of density waves. In gases of ultracold atoms the critical velocity has been measured experimentally both for bosons \cite{Raman99Critvel} and also for fermions \cite{Miller07Critvel}. Though some doubt due to the inhomogeneity of the condensate and the finite size of the disturbing potential remains, the experiments confirm a speed of sound around that of the Landau criterion. Recently Navez and Graham showed for a spinless Bose gas in a dynamically consistent Hartree--Fock-RPA approximation \cite{Navez06Critical} that at the critical velocity not only the quasiparticle energy becomes zero but also the damping rate changes sign signaling the onset of a dynamical instability.

Here we show for antiferromagnetic spin-1 Bose gases that the critical velocity is the speed of spin-wave instead of the sound wave. In typical experiments with ultracold gases the spin waves are usually two orders of magnitude slower than sound waves. Therefore the critical velocity is much smaller than in a scalar gas. A direct physical consequence is that when a spinor Bose condensate moves faster than the speed of the spin wave, it emits a cone of spin polarized excitations instead of density waves, which can lead to spin texture formation and can be measured in experiments.

\section{The formulation of the problem of an antiferromagnetic spin-1 Bose gas with a moving condensate}
\label{sec:form}

We consider an ultracold, dilute, homogeneous, spin-1 Bose gas  in a homogeneous magnetic field. The interparticle
interaction is modeled by s-wave scattering, i.e. we neglect the relatively weak dipolar interaction of the gas. The grand-canonical Hamiltonian of the system reads as
\begin{multline}
\label{eq:ham}
  {\mathcal H}=\sum_{\genfrac{}{}{0pt}{2}{\vec{k}}{r,s}}
  \Big[(e_{\vec{k}}-\mu)\delta_{rs} -g \mb B\,
  (F_z)_{rs}\Big] a_r^\dagger(\vec{k})
  a_s(\vec{k})\\+\frac{1}{2V}\sum_{\genfrac{}{}{0pt}{2}{\vec{k}_1+
    \vec{k}_2=\vec{k}_3+\vec{k}_4}{r,s,r',s'}}a^\dagger_{r'}(\vec{k}_1)
  a^\dagger_r(\vec{k}_2)V^{r's'}_{rs}a_s(\vec{k}_3)a_{s'}
  (\vec{k}_4),
\end{multline}
where $a_r(\vec{k})$ is the annihilation operator of plane wave states with momentum $\vec{k}$ and spin projection
$r$. The spin index $r$ refers to the eigenvalue of the \z-component of the
spin operator and can take values from $+,0,-$. Correspondingly $F_z=\mathrm{diag}(1,0,-1)$ is a 3x3 diagonal matrix. In Eq. \eqref{eq:ham} $\kink=\hslash^2k^2/(2M)$ refers to the kinetic energy of an atom, $\mu$ to the chemical potential, $g$ to the gyromagnetic ratio, $\mu_{\mathrm{B}}$ to the Bohr magneton, $B$ to the modulus of the homogeneous magnetic field. $V$ is the volume of the system and $V^{r's'}_{rs}$ the amplitude of the two particle interaction, given for spin-1 bosons by \cite{Ho98Spinor,Ohmi98Spinor,Stamper-Kurn2001Lecture}:
\begin{equation}
  \label{eq:pseudopot}
    V^{r's'}_{rs}=c_n\delta_{rs}\delta_{r's'}+c_s(\vec{F})_{rs}
    (\vec{F})_{r's'},
\end{equation}
with $c_n=4\pi\hslash^2(a_0+2a_2)/(3M)$ , and $c_s=4\pi\hslash^2(a_2-a_0)/(3M)$.
The parameters $a_0$ and $a_2$ are the scattering lengths in the total hyperfine
spin channel zero and two, respectively. The constant $c_n>0$, while $c_s$ can
both be positive or negative, depending on the relative values of $a_0$ and $a_2$.
For $c_s>0$ a zero net spin is energetically favorable (in the absence of a magnetic field). For
this reason systems with $c_s>0$ are called antiferromagnetic or polar gases \cite{Ho98Spinor}. For instance the ultracold
gas of  \Na in the $f=1$ hyperfine state is antiferromagnetic \cite{Crubellier99Na}.

Even in the presence of a homogeneous magnetic field the \z\ component of the total spin is a conserved quantity. This conservation law can be resolved similarly as that for the particle number [i.e., with the introduction of a Lagrange multiplier in the Hamiltonian \eqref{eq:ham}]. This multiplier shows up in the same way as the magnetic field does, therefore an effective magnetic field can be introduced as a sum of the external magnetic field plus the Lagrange multiplier. In the following, $B$ will mean this kind of effective field even when the physical magnetic field is zero. In the following, we assume that either such is the case, or the external magnetic field is so small that the quadratic Zeeman effect can be neglected.

In the presence of a Bose-Einstein condensate moving with velocity $\vec{v_0}=\hslash \vec{k}_0/M$ the annihilation and creation operators have a nonzero mean value $a_r(\vec{k}_0)=\sqrt{N_c}\zeta_r$, where $N_c$ is the number of atoms in the condensate, and $\zeta_r=(\zeta_+,\zeta_0,\zeta_-)$ is the condensate spinor normalized to unity \cite{Ho98Spinor}. It is convenient to define new annihilation operators with the canonical transformation $b_r(\vec{k})=a_r(\vec{k})-\sqrt{N_c}\zeta_r\delta_{\vec{k},\vec{k}_0}$ which have zero mean values for all $\vec{k}$. The finite temperature Green's function of the system is introduced as
\begin{equation}
\label{eq:greensfn}
\mathcal{G}^{rs}_{\alpha\beta}(\vec{k},\tau)=-\left\langle T_\tau b^\alpha_r(\vec{k},\tau)b^{\beta\dagger}_s(\vec{k},0)\right\rangle,
\end{equation}
where $\tau$ is the imaginary time, $T_\tau$ is the $\tau$ ordering operator and the Greek indices take the values $\pm1$ and are introduced for a shorthand notation, namely $b^{1}_r(\vec{k},\tau)=b_r(\vec{k},\tau)$ and $b^{-1}_r(\vec{k},\tau)=b^\dagger_r(2\vec{k}_0-\vec{k},-\tau)$. The Green's function can be conveniently arranged into a matrix form
\begin{equation}
\label{eq:grmat}
\tens{G}=\left[
\begin{array}{l r|l r}
\mathcal{G}^{++}_{11}&\mathcal{G}^{++}_{1,-1}&\mathcal{G}^{+-}_{11}&\mathcal{G}^{+-}_{1,-1}\\
\mathcal{G}^{++}_{-1,1}&\mathcal{G}^{++}_{-1,-1}&\mathcal{G}^{+-}_{-1,1}&\mathcal{G}^{+-}_{-1,-1}\\
\hline
\mathcal{G}^{-+}_{11}&\mathcal{G}^{-+}_{1,-1}&\mathcal{G}^{--}_{11}&\mathcal{G}^{--}_{1,-1}\\
\mathcal{G}^{-+}_{-1,1}&\mathcal{G}^{-+}_{-1,-1}&\mathcal{G}^{--}_{-1,1}&\mathcal{G}^{--}_{-1,-1}
\end{array}
\right].
\end{equation}
In Matsubara representation the inverse of the Green's function is expressed from the Dyson-Beliaev equations with the help of the free propagator and the self-energy
\begin{equation}
\label{eq:dysonbel}
\tens{G}^{-1}(\vec{k},i\omega_n)=\tens{G}^{-1}_0(\vec{k},i\omega_n)-\tens{\Sigma}(\vec{k},i\omega_n).
\end{equation}
The self-energy matrix $\tens{\Sigma}$ is arranged from its components in a similar way than the Green's function \eqref{eq:grmat}.
The inverse of the free propagator is given by the diagonal matrix
\begin{equation}
\label{eq:freeprop}
\tens{G}^{-1}_0(\vec{k},i\omega_n)=\left[
\begin{array}{l c c r}
\Xi+\omega_L&0&0&0\\
0&\tilde\Xi+\omega_L&0&0\\
0&0&\Xi-\omega_L&0\\
0&0&0&\tilde\Xi-\omega_L
\end{array}
\right],
\end{equation}
with $\Xi=i\omega_n-\hslash^{-1}(\kink-\mu)$, $\tilde\Xi=-i\omega_n-\hslash^{-1}(\kinkt-\mu)$ and $\omega_L=\hslash^{-1}g\mb B$ the Larmor frequency related to the effective magnetic field. The self-energy $\tens{\Sigma}(\vec{k},i\omega_n)$ is to be chosen according to some approximation scheme. The dispersion relation $\omega(\vec{k})$ of the quasiparticles is provided by the poles of the retarded Green's function, that is
\begin{equation}
\label{eq:disprel}
\det\tens{G}^{-1}(\vec{k},\omega+i\eta)=0.
\end{equation}
To obtain the retarded Green's function we analytically continue in the frequency variable $i\omega_n\rightarrow\omega+i\eta$ (with $\eta\rightarrow +0$) through the upper half complex plane. In the following sections we are going to evaluate the self-energy first in the Bogoliubov approximation valid  at zero temperature, and later in the Hartree-RPA approximation in order to study also the temperature dependence of the critical velocity.
Though we investigate the Green's functions it is important that they are coupled to the propagators describing collective motion and as a result their excitations hybridize \cite{Szepfalusy01Structure}.

\section{Bogoliubov theory of a spin-1 Bose gas with a moving condensate}
\label{sec:bogo}

The simplest possible approximation is the Bogoliubov approximation, which is valid at zero temperature or very close to it. In the Bogoliubov approximation the noncondensed atom fraction is assumed to be negligible, therefore the condensate atom number is approximately equal to the total number of atoms: $N_c\approx N$.

There are two distinct phases distinguished by the number of condensate components, which we call as P1 and P2 phases in analogy to the case of superfluid ${}^3\mathrm{He}$. When the gas sample is maximally polarized all of the condensate atoms are in a single spin component and the condensate spinor is $\zeta_r^{\mathrm{P1}}=(1,0,0)$ \cite{Kis-Szabo07Polar,Szirmai12Hydro}. In spite of the antiferromagnetic coupling the condensate is maximally magnetized due to the initial preparation and the conservation of the magnetization. When the polarization of the gas is decreased the condensate wavefunction becomes multicomponent $\zeta_r^{\mathrm{P2}}=(\zeta_+,\zeta_0,\zeta_-)$. Note that the component $\zeta_0=0$ always in the Bogoliubov approximation, which is the consequence that the magnetization of the condensate points in the direction of the magnetic field (although our case it is just a Lagrange multiplier). Consequently the \x\ and \y\ components of the condensate magnetization have to vanish and therefore $\zeta_0=0$ \cite{Szirmai12Hydro}. In this paper we restrict the discussion to the P2 phase.

\begin{figure}[tb!]
\centering
\includegraphics{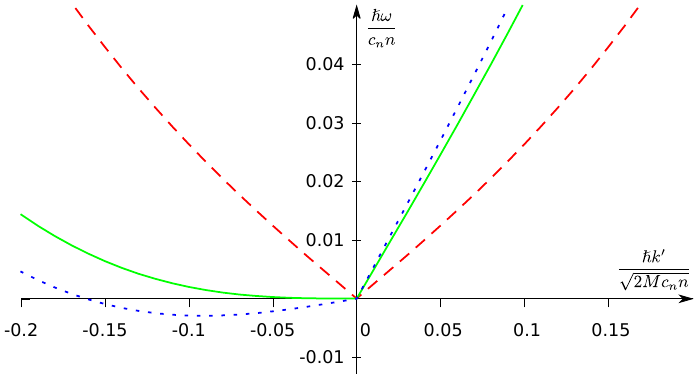}
\caption{Dispersion relation of spin waves in the Bogoliubov approximation. The dashed (red) line is for $v_0=0$, the solid (green) line is for $v_0=v_{0,c}=c_2$, the dotted (blue) line is for $v_0=1.2\times v_{0,c}$. The parameters are $T=0$, $\hslash \omega_L=0.1 c_s n$, and $c_s=3\times10^{-2}c_n$.}
\label{fig:bogodisp}
\end{figure}
The equation of state follows from the generalized Hugenholtz-Pines theorem \cite{Kis-Szabo07Polar}. When the condensate is not fully polarized, that is when $\zeta_-\neq0$, we have two separate equations, one for the chemical potential and one for the magnetic field.
\begin{subequations}
\label{eqs:eqstate}
\begin{align}
\mu&=e_{\vec{k}_0}+n c_n,\\
\hslash\omega_L&= m c_s,
\end{align}
\end{subequations}
where $n=n_c=n_{c,+}+n_{c,-}$ is the total denstiy of atoms, and $m=n_{c,+}-n_{c,-}$ is the total magnetization density. The condensate density in spin component $r$ is given by $n_{c,r}=n_c\,\zeta_r$. Equations \eqref{eqs:eqstate} relate the chemical potential ($\mu$), the magnetic field ($\hslash\omega_L$) to the density $n$, the magnetization $m$ and the velocity of the fluid ($\hslash \vec{k}_0/M$).

The self-energy in the Bogoliubov approximation is given by
\begin{equation}
\label{eq:slefenergB}
\Sigma^{rs}_{\alpha\beta}(\vec{k},i\omega_n)=\hslash^{-1}\big[(n\,c_n+r\,m\,c_s)\delta_{\alpha\beta}\delta_{rs}+\sqrt{n_{c,r}\,n_{c,s}}(c_n+r\,s\,c_s)\big].
\end{equation}
Combining Eqs. \eqref{eq:dysonbel}, \eqref{eq:freeprop}, \eqref{eq:disprel} and \eqref{eq:slefenergB} and performing some straightforward algebra one obtains the following equations for the dispersion relation of the quasiparticles:
\begin{subequations}
\label{eq:dispreleqB}
\begin{align}
0&=\rho^2-\rho(c_n+c_s)n+4 c_n c_s n_{c,+} n_{c,-},\label{eq:dispreleq1}\\
 \rho\Big(\Delta e+\widetilde{\Delta e}\Big)&=\Big(\omega-\Delta e\Big)\Big(\omega+\widetilde{\Delta e}\Big).\label{eq:dispreleq2}
\end{align}
\end{subequations}
Where we have introduced the quantity $\rho$, which is provided by the solution of Eq. \eqref{eq:dispreleq1}. We have also introduced $\Delta e=\kink-e_{\vec{k}_0}$ and $\widetilde{\Delta e}=\kinkt-e_{\vec{k}_0}$. By measuring the wave number relative to the condensate $\vec{k}'=\vec{k}-\vec{k}_0$, the solution of Eq. \eqref{eq:dispreleq2} is
\begin{subequations}
\label{eqs:bogodisprelfinal}
\begin{equation}
\label{eq:dispB}
\omega_{\vec{k}'}=\vec{k}'\cdot\vec{v}_0\pm\hslash^{-1}\sqrt{e_{\vec{k}'}^2+2\rho e_{\vec{k}'}},
\end{equation}
from which we see that the speed of the quasiparticle for the condensate in rest, i.e. the proportionality constant in the starting linear part of the dispersion relation, is simply  $c_{1,2}=\sqrt{\rho_{1,2}/M}$. The quantity $\rho$ is obtained from the quadratic equation \eqref{eq:dispreleq1}:
\begin{equation}
\label{eq:rho}
\rho_{1,2}=\frac{(c_n+c_s)n}{2}\pm\frac{1}{2}\sqrt{(c_n-c_s)^2 n^2+4 c_n c_s m^2}.
\end{equation}
\end{subequations}

It is easy to see that both solutions \eqref{eq:rho} are positive. The dispersion relation Eqs. \eqref{eqs:bogodisprelfinal} for $\vec{v}_0=0$ are in complete agreement with the results of Ohmi and Machida \cite{Ohmi98Spinor}. The two solutions correspond to the density wave (with the + sign) and spin wave (with the - sign) excitations. When $c_s\ll c_n$ the speed of the spin waves is proportional to $(c_s)^{1/2}$, while the speed of sound is proportional to $(c_n)^{1/2}$.

The dispersion relation \eqref{eq:dispB} is very intuitive in accordance with Landau's idea: the first term in Eq. \eqref{eq:dispB} is just what we expect after a Galilean transformation to the frame moving with the condensate, while the second therm coincides with the result in the frame where the condensate is at rest.

When the condensate moves with velocity $\vec{v}_0$, excitations with wavenumber parallel to the condensate have a steeper dispersion while those which move against the condensate have a flatter one. The critical velocity $\vec{v}_{0,c}$ is where the dispersion of the oppositely moving quasiparticles is horizontal. Since the spin wave velocity is smaller than the speed of sound, the instability due to the motion of the condensate appears in the spin wave dispersion first. Therefore the critical velocity is the speed of spin wave for a condensate at rest: $v_{0,c}=c_2$. In Fig. \ref{fig:bogodisp} we show the spin wave dispersion for the three cases.  

\section{Mean-field approximation at finite temperature}
\label{sec:hartree}

\begin{figure}[tb!]
\centering
\includegraphics{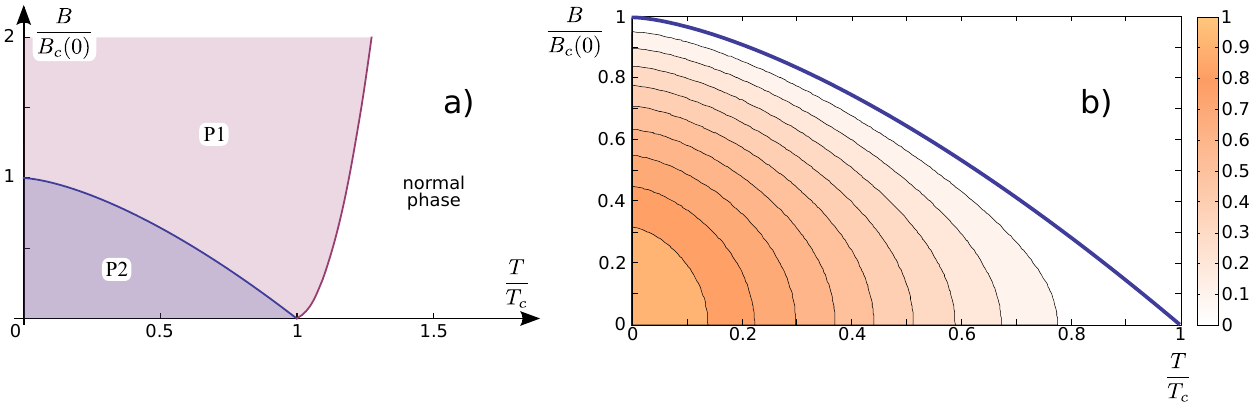}
\caption{a) The phase diagram in the Bogoliubov-Hartree approxmiation. $B_c(0)=c_s n/(g\mb)$ is the critical value of the magnetic field separating the P1 and P2 phases at $T=0$. b) the dependence of the square of the critical velocity (in units of $c_s n/M$) on the temperature and magnetic field in the P2 phase.}
\label{fig:hartree}
\end{figure}

In order to evaluate both the temperature and magnetization dependence of the critical velocity we use the Landau criterion generalized to the spinor case according to the previous section. Here we use the finite temperature excitation spectrum of a spinor Bose gas in the P2 phase, worked out earlier in \cite{Kis-Szabo07Polar}, and evaluate the speed of spin wave of a condensate at rest, i.e. the critical velocity, at different temperature regions. At finite temperatures the condensate densities depend both on temperature and magnetic field. The partially polarized P2 phase occupies a finite region in the space of temperature and magnetic field (or rather magnetization). When the magnetization is zero the P2 phase goes directly to the noncondensed phase. However, when the gas sample is polarized the P2 phase first goes to a diefferent BEC phase, the P1 phase, where the condensate spinor is single component. The phase diagram is depicted in Fig. \ref{fig:hartree} (a). This approximation takes into account the effect of thermal excitations in a dynamically consistent way: this approximation is conserving and gapless and also accounts for the hybridization of one-particle and collective excitations \cite{Szepfalusy01Structure}. Note that to ensure the coincidence of one particle and collective motions already in the Hartree approximation one needs to sum up bubble diagrams \cite{Szepfalusy01Structure}.

%
%
For further discussions we consider the experimentally relevant situation, when $c_s\ll c_n$. In fact, for \Na the ratio of coupling constants is $\epsilon\equiv c_s/c_n\approx3\times10^{-2}$.
Besides the thermal wavelength, we can introduce two other characteristic lengths: $\xi_B=\hslash/(4Mc_sn_{c,-})^{1/2}$ is the mean-field coherence length, and $\xi'=Mk_BT/(4\pi\hslash^2n_{c,-})$ is the characteristic length governing the phase fluctuations of the order parameter.  As discussed in \cite{Kis-Szabo07Polar}, at  sufficiently low temperature, where the longest characteristic length is the thermal wavelength $\Lambda\gg\xi_B,\xi'$, 
the excitation energies are linear in wavenumber: $\omega_{1,2}=\sqrt{\rho_{1,2}/M}k'$, where $\rho_1=n (c_n+c_s)-4c_sn_+n_-/n$ is for conventional sound waves, and $\rho_2=n_0c_s+4c_s n_+n_-/n$ describes spin waves. The zero temperature limit agrees with the the results \eqref{eq:rho} and with \cite{Ohmi98Spinor}, however here $n_r$ is the total particle density in spin component $r=\lbrace0,+,-\rbrace$ coming from the condensate contribution $n_{c,r}$ and the thermal one $n'_r$. Note that $n_0=n_0'$ is entirely thermal. It is remarkable that at medium temperatures, characterized by $\xi_b\ll\xi',\Lambda$, the square of the spin wave velocity becomes
\begin{equation}
\label{eq:spinvelH2}
c_2^2=\frac{\rho_2}{M}=\frac{4n_{c,+}n_{c,-}}{n}\frac{c_s}{M},
\end{equation}
but now an essential difference is that the densities appearing in Eq. \eqref{eq:spinvelH2} are those of the condensate, that is they exhibit significant temperature dependence. We do not discuss the critical region where the mean-field theory breaks down, anyhow, one expects that the critical velocity tends to zero there. Fig \ref{fig:hartree} (b) shows the temperature and magnetic field dependence of the square of the critical velocity $c_2$. The critical velocity vanishes at the P1-P2 phase boundary but even in the P2 phase its value can be very small when $n_{c,-}$ is small.

We have discussed the problem of critical flow in spin-1 Bose gases in the P2 phase. We have shown that the critical velocity is the speed of spin waves and is much smaller than the critical velocity in scalar gases. We have not dicussed the P1 phase, but one can expect, that there, close to the P1-P2 phase boundary the gap of the spin wave plays the role of critical velocity.

\begin{acknowledgement}
It is my pleasure to thank P\'eter Sz\'epfalusy for many useful discussions. This work was supported by the Hungarian National Research Fund (OTKA T077629). Support from the Hungarian Academy of Sciences (Lend\"ulet Program, LP2011-016), from the Hungarian National Office for Research and
Technology (ERC\_HU\_09 OPTOMECH) and from the J\'anos Bolyai Scholarship is also kindly acknowledged.
\end{acknowledgement}

\end{document}